\documentclass[11pt,a4paper]{article}
\usepackage{marvosym}
\usepackage{a4wide}
\usepackage{amsthm}
\usepackage{amsmath}
\usepackage{amssymb}
\usepackage{natbib,epsfig}

\def\fixorder#1{}

\def\marginpar#1{}
\def\endnote#1{}{}

\theoremstyle{plain}

\newtheorem{thm}{Theorem}

\newtheorem{lemma}{Lemma}

\newtheorem{alg}{Algorithm}

\theoremstyle{definition}

\title{Network Tomography: Identifiability and Fourier Domain Estimation}
\author{Aiyou Chen, Jin Cao and Tian Bu\\
(aychen,cao,tbu)@research.bell-labs.com\\
Bell Laboratories, Alcatel-Lucent Technologies\\
Murray Hill, NJ 07974, USA
}

\begin{document}
    
\maketitle

\begin{abstract}

The statistical problem for network tomography is to infer the
distribution of 
$\mathbf{X}$, with mutually independent components, from a measurement
model $\mathbf{Y}=A\mathbf{X}$, 
where $A$ is a given binary matrix representing the routing topology of a
network under consideration.
The challenge is that the dimension of $\mathbf{X}$ is  
much larger than that of $\mathbf{Y}$ and thus the problem is
often called ill-posed. This paper studies some statistical
aspects of network tomography. We first address the identifiability issue
and prove that the $\mathbf{X}$ distribution is identifiable up to a
shift parameter under mild conditions. We then use a mixture model of
characteristic functions to derive a fast algorithm for estimating the
distribution of $\mathbf{X}$ based on the General method of Moments.
Through extensive model simulation and real Internet trace driven
simulation, the proposed approach is shown to be favorable
comparing to previous methods using simple discretization
for inferring link delays in a heterogeneous network.


\end{abstract}

\section{Introduction}

Network performance monitoring and diagnosis is challenging due to the
size and decentralized nature of the Internet.
The service providers may collect their link level
statistics using tools such as Cisco Netflow, whereas the end users
can obtain the end-to-end performance by probing the
network. Unfortunately, none of them has a global view of the
Internet. For instance, when an end-to-end measurement indicates the
performance degradation of an Internet path, the exact cause is hard
to be uncovered because the path may traverse several autonomous
systems (AS) that are often owned by different entities and the
service providers generally do not share 
their internal performance.  Even if they do, there is no scalable way to
correlate the link level measurements to end-to-end performance in a
large network like the Internet. Similarly, the service providers may
be interested in the end-to-end path characteristics that they can not
observe directly. 
Network tomography is a technology addressing these issues that infers
unobservable characteristics from easily available measurements.
There have been two forms of network tomography being studied in the
literature. One, called {\it network delay tomography},
estimates the link-level characteristics based on
end-to-end measurements, 
and the other, called {\it traffic demand tomography}, predicts
end-to-end path-level traffic intensities based on link-level traffic
measurements. 
The key advantage of network tomography is that it does not require
the collaboration between network internal elements and end users.
See \cite{castro:2004}, \cite{denby:2007} and references
therein for an excellent review.
We focus on network delay tomography in this paper, while the proposed
approach may also be applied to traffic demand tomography.

Network delay tomography aims to estimating network internal
characteristics such as loss and delay\footnote{To be precise, the
  delay here is the queuing delay that excludes the constant link
  propagation delay, we omit queuing when context is clear}, from 
end-to-end measurements by exploiting the inherent correlation in
performance.  Considering a tree spanning a source of probes (root)
and a set of receivers (leaves), the packets are potentially subject
to queuing delay and loss at each link. 
The end-to-end (source-to-receiver) measurements may be made passively
or actively. The  
probes for the active measurements can be sent using either multicast
or unicast routing\footnote{With multicast, a packet is sent from a
  source to multiple destinations simultaneously; with unicast, a
  packet is sent to different destinations separately}. See 
\cite{lawrence.et.al.2006} and \cite{denby:2007} for examples of how
unicast and multicast 
probes can be designed and sent. Because only
one copy of a probe is transmitted  
on the common links, multicast probing based tomography has the
advantage of perfect correlation on the common links, less
overhead, and better scalability. Following \cite{presti:2002} and
\cite{liang:2003}, we assume that 
measurements are collected from multicast probes, although the multicast
routing is not widely enabled in today's Internet. It has been shown in
\cite{bu:2002} on how to apply the tomography algorithms developed for
multicast measurements when only unicast measurements are available.

The statistical models for both types of network tomography can be
unified as follows:
\begin{equation}\label{eq:tomo1}
\mathbf{Y}=A\mathbf{X}, 
\end{equation} 
where
$\mathbf{X}=(X_{1},\ldots,X_{J})^T$ is a $J$-dimensional vector of
network dynamic parameters, and $\mathbf{Y}=(Y_{1},\ldots,Y_{I})^T$ is
an $I$-dimensional vector of measurements and $A$ is an $I\times J$
matrix with elements 0 or 1 which represents the routing topology of
the network under 
consideration. Here we use the superscript $T$
to denote the transpose. In most network tomography scenarios, the
components of $\mathbf{X}$ are assumed independent but 
unobservable. Usually $I$ can be as large as $J^2$ for network demand
tomography and as large as $2J-1$ for network delay tomography. In
network delay tomography, each component of $\mathbf{X}$ represents an
internal link delay and each component of $\mathbf{Y}$ represents a
delay measurement from a source to a destination. 
The objective of network tomography is to estimate the distribution of 
$\mathbf{X}$ given independent observations from the distribution of
$\mathbf{Y}$.

\begin{figure}[t]
\begin{minipage}[t]{.42\textwidth}
\begin{center}         
\epsfig{file=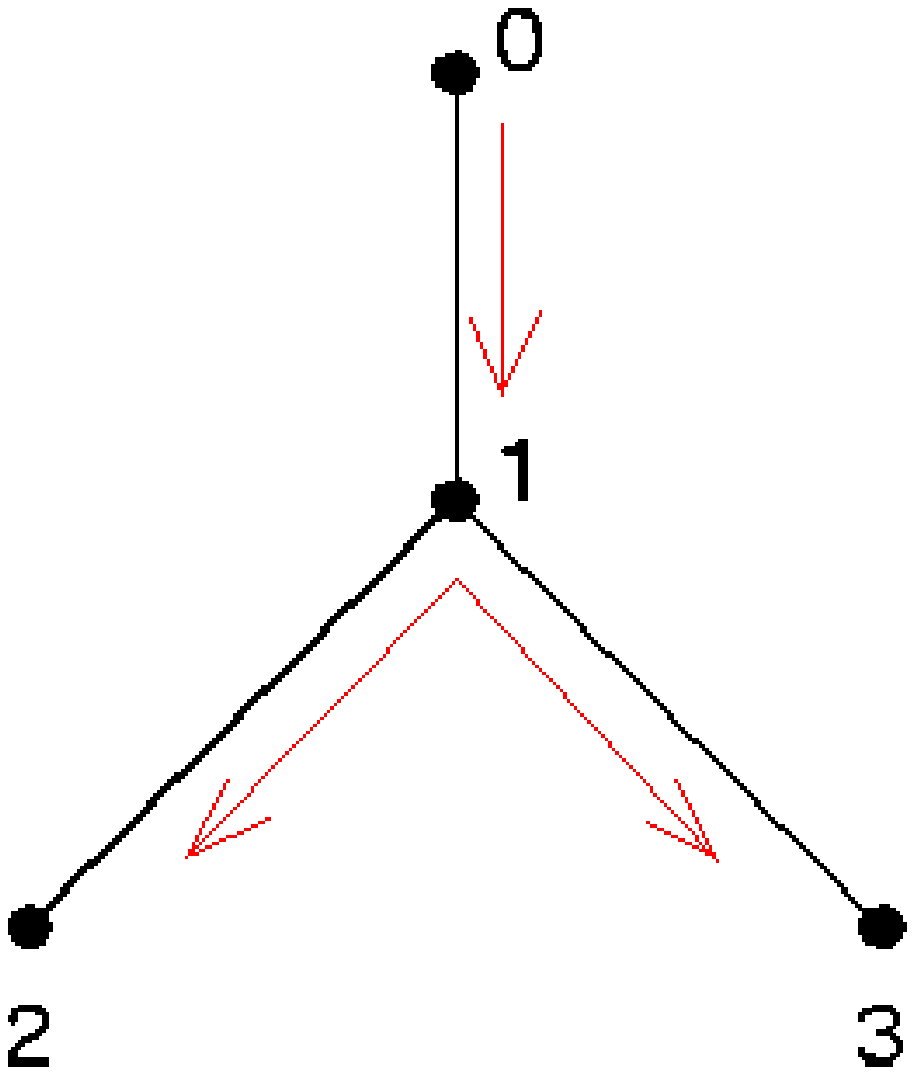, scale=0.35}       
\caption{Two-leaf tree}       
\label{fig:twoleaf}     
\end{center}   
\end{minipage}
\begin{minipage}[t]{.42\textwidth}     
\begin{center}
\epsfig{file=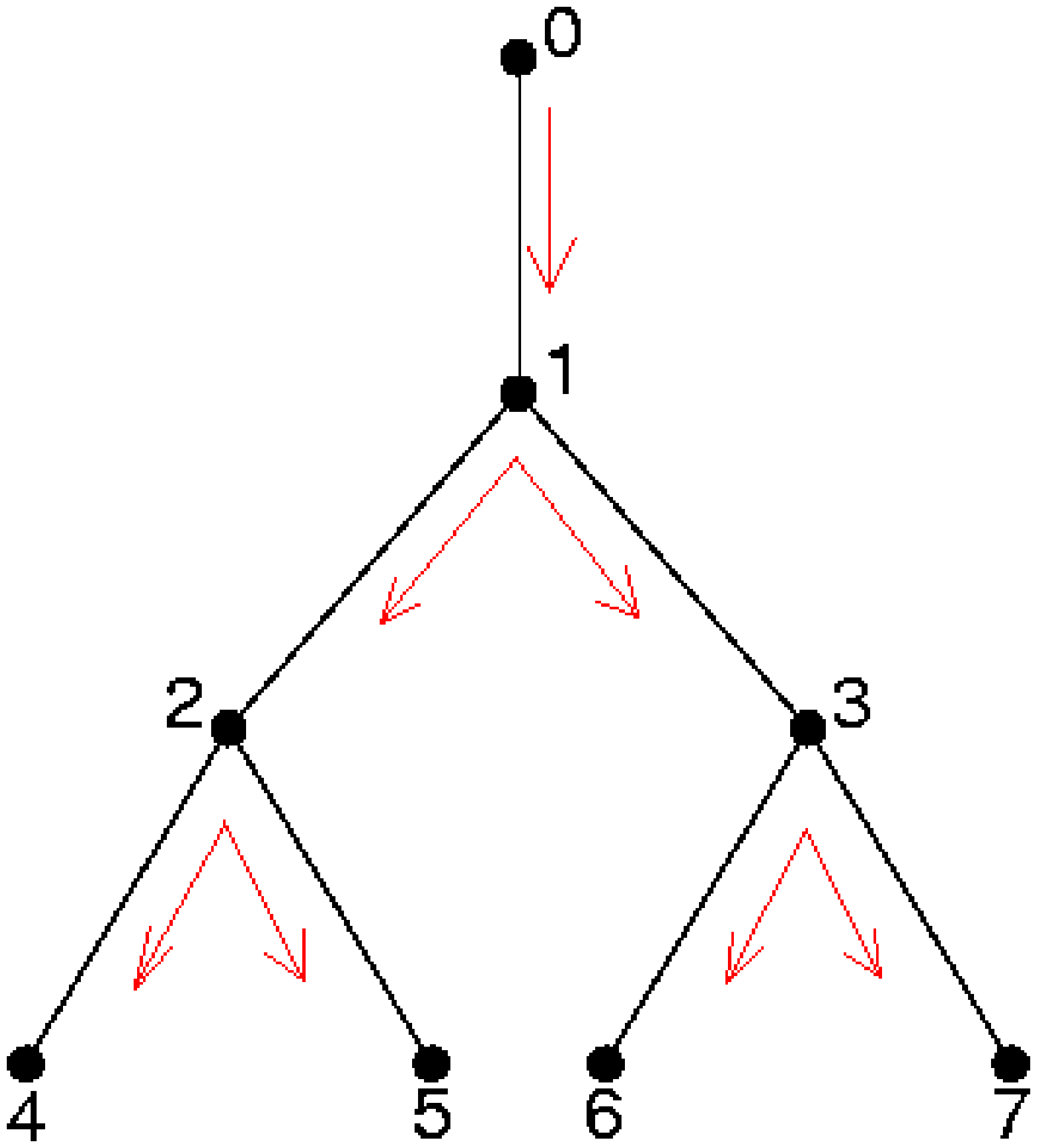, scale=0.3}       
\caption{Four-leaf tree}
\label{fig:4leaf}     
\end{center}
\end{minipage}
\end{figure}

As a simple example,
Figure \ref{fig:twoleaf} shows a two-leaf tree topology, on which a
probing packet is sent from the root node 0 (source) to leaf nodes 2
and 3 (receivers). When the packet arrives at node 1 from the source, it is
replicated and transmitted to node 2 and 3
simultaneously as the red arrows show. Let $X_1$ denote the link delay
from node 0 to node 1 and 
let $X_i$ denote the link delay from node 1 to node $i$, $i=2,3$
respectively. Let $Y_1, Y_2$ be the end-to-end delays from node 0 to
node 2 and 3 respectively. Then $Y_1=X_1+X_2$ and $Y_2=X_1+X_3$, which
can be written in the form of \eqref{eq:tomo1} with $A$ a $2\times 3$
binary matrix, i.e. $A=[1,1,0;1,0,1]$.

There have been significant amount of works on network tomography in
recent years. Network tomography was first proposed by
\cite{vardi:1996} and then followed by remarkably \cite{tebaldi:1998}, 
\cite{cao:2000a} and \cite{liang:2003} 
for traffic matrix estimation, i.e. traffic demand tomography. 
\cite{caceresIT:1999}, \cite{zhu.geng.2005} and \cite{Xi.et.al.2006} among others 
studied it for inferring network internal loss.
Network delay tomography has also been studied
extensively. \cite{presti:2002} developed a fast algebraic
algorithm but it is quite inefficient. \cite{bu:2002} showed that
the maximum likelihood estimate (MLE) requires exponential computational
complexity. \cite{tsang:2003} proposed a penalized maximum
likelihood method.  \cite{liang:2003} proposed a pseudo-likelihood method with
multicast measurements, and recently \cite{lawrence.et.al.2006} proposed
local likelihood method with both unicast and multicast measurements,
both of which were shown to be fast and quite efficient compared with the MLE.
These studies are based on a discrete distribution with equally
spaced bins for modeling link delays, where the {\it same} 
bin width is used for all the links for the ease of
computation. 

\cite{duffield:2001b} pointed out that, however, a single fixed bin
 width is not appropriate for heterogeneous networks such as the
 Internet because it does not scale well between both fast links and
 slow links.
They proposed a varying-bin discrete model for estimating link delay
distributions based on unicast measurements. Their estimation idea is to use
structured bins such that they can iteratively estimate a segment 
of delay distributions by truncating the delays from both sides,
i.e. rounding the left of the segment to zero and the right to
infinity. However, 
the performance of their estimation approach is not better than that 
using an equal-bin discrete model with an appropriate bin width
as they reported, probably due to the 
bias introduced by their brute-force truncation. Our approach is
also based on varying-bin type models but does not suffer from such bias.
\cite{shih.hero.2001} proposed to estimate cumulative 
generating functions (similar to characteristic functions used in this
paper) of link delays, but 
they did not estimate link delay distributions.  \cite{shih.hero:2003}
also proposed finite mixture models with Gaussian components for link delay
distributions based on unicast measurements.

There are several previous works that have considered the identifiability issue
for the network tomography problem, for example \cite{vardi:1996},
\cite{cao:2000a} and \cite{presti:2002}. These authors considered
instances of the tomography problem by assuming specific parametric
(such as Poisson and Gaussian) or discrete distributions.
We will unify these results and extend the identifiability
condition to general distributions under mild assumptions.

The contributions of this paper are as follows. 
First, we prove that the
distribution of $\mathbf{X}$ is identifiable up to a shift parameter
under general conditions. 
Second, we
propose flexible mixture models of characteristic  
functions for network delay
tomography and develop a fast algorithm 
for estimation based on the General Method of Moments (GMM).
The new approach allows one to model continuous
delays on {\it heterogeneous} network links conveniently, where delays
may not have the same scale across all network links. 
Extensive model simulation and real Internet trace-driven
simulation suggest that our new approach can yield more accurate
estimates of link delay distributions yet is computationally less
expensive than previous approaches. 

The remaining sections of the paper are structured as follows.
In Section 2, we address the 
identifiability issue. We describe the mixture models for link delays
in Section 3 and develop a fast algorithm for estimating the delay
distributions in Section 4.  In Section 5, we present 
extensive experimental studies for evaluating the proposed method.  Section 6
concludes the paper.

\section{Identifiability}

In this section, we study the identifiability issue for model
 \eqref{eq:tomo1} and prove that the distribution of $\mathbf{X}$ is
 identifiable up to a shift parameter under mild conditions. The main
 tool we use is characteristic function whose basic properties are
 reviewed below.

\subsection{Characteristic Function}
A characteristic function of a univariate random variable $Z$ is defined by
\[
\phi_{Z}(t)=E[e^{itZ}]=\int_{-\infty}^{\infty}e^{itz}f_{Z}(z)dz,\,\,\,\,
t\in\mathcal{R},\] where $E[\cdot]$ denotes the expectation with
respect to $Z$ and
$f_{Z}(\cdot)$ is the probability density function of $Z$. By
convention, $\phi_Z$ and $f_Z$ denote the characteristic function and
probability density function of $Z$, respectively.
The characteristic function for a random vector
$\mathbf{Z}\in\mathcal{R}^D$ can be defined in a similar 
manner by considering $\mathbf{t}\in \mathcal{R}^D$ instead of
$\mathcal{R}$. It is well known that a probability distribution can be
uniquely specified by its characteristic function and vise versa. 

Suppose $Z_{1}$ and
$Z_{2}$ are two independent random variables. Then the joint
characteristic function of $\mathbf{Z}=(Z_1,Z_2)$ is 
\[
\phi_\mathbf{Z}(\mathbf{t}) = \phi_{Z_1}(t_1) \phi_{Z_2}(t_2), \ \
\mathbf{t}=(t_1,t_2), 
\]
which is a product of the marginal characteristic functions of $\mathbf{Z}$. Let
$V=Z_1+Z_2$, then the characteristic function of $V$ is simply a
product of the characteristic functions of $Z_{1}$ and $Z_{2}$, i.e.,
\[
\phi_{V}(t)=\phi_{Z_{1}}(t)\phi_{Z_{2}}(t).
\]
This is much easier to compute than the density function
of $V$, say $f_{V}(\cdot)$, which is 
a convolution of densities of $Z_{1}$ and $Z_{2}$, i.e.,
\[
f_{V}(v)=\int_{z_{1}\in\mathcal{R}}f_{Z_{1}}(z_{1})f_{Z_{2}}(v-z_{1})dz_{1}.
\]
For the tomography model \eqref{eq:tomo1}, since the components of
$\mathbf{X}$ are mutually independent,
it is easy to evaluate the characteristic function of $\mathbf{Y}$ by
\begin{eqnarray}\label{eq:CFY}
\phi_{\mathbf{Y}}(\mathbf{t}) = E[e^{i\mathbf{t}^{T}\mathbf{Y}}]  =
E[e^{i(\mathbf{t}^{T}A)\mathbf{X}}] = 
\prod_{j=1}^{J}\phi_{X_{j}}(\mathbf{t}^{T}A^{j}),
\end{eqnarray}
where $A^j$ is the $j$th column of $A$. However, it is in general
difficult to evaluate the distribution of $\mathbf{Y}$ because it is a
high order convolution in terms of the distribution of $\mathbf{X}$.
Below we will use the formula \eqref{eq:CFY} for both the identifiability
proof and estimation in network tomography.

\subsection{Identifiability}
By identifiability, we mean that the distribution of $\mathbf{X}$ can be
uniquely determined by the distribution of $\mathbf{Y}$. It is
important to establish the identifiability. Otherwise, the distribution
of $\mathbf{X}$ may not be estimable from the distribution of
$\mathbf{Y}$.  In the following, we present our general theorems for
identifiability and discuss related issues.

We assume that $E|X_j|<\infty$, $j=1,\cdots,J$ and that the distribution of $\mathbf{X}$ satisfies one of the
 two conditions, namely C1 and C2, defined below.
\begin{itemize}
\item[(C1)] the characteristic function of each $X_j$  is analytic \footnote{An analytic characteristic function corresponds to a distribution function which has moments $m_k$ of all orders $k$ and 
$\lim\sup_{k\rightarrow\infty}[|m_{k}|/k!]^{1/k}$ is finite}; 
\item[(C2)] the characteristic function of each $X_j$ has no zeros in $\mathcal{R}$. 
\end{itemize}

We first address the identifiability issue in Lemma 1 for the
simple two-leaf 
tree tomography model described earlier with Figure
\ref{fig:twoleaf}. The result will serve as the basis for Theorem 1
and 2 below where the routing topology is not a simple two-leaf tree.

\begin{lemma}\label{lemma:lemma}
If  \(Y_{1}=X_{1}+X_{2}\) and \(Y_{2}=X_{1}+X_{3}\), then the distributions of $X_1,X_2,X_3$ can be identified up to a shift parameter.
\end{lemma}

\begin{proof}
Suppose there exist both $\mathbf{X}=(X_1,X_2,X_3)^T$ and $\mathbf{X}'=(X_1',X_2',X_3')^T$ with mutually independent
components that give rise to the same distribution $\mathbf{Y}=(Y_1,Y_2)$, then we show
that distributions of $X_j$ and $X_j'$, $j=1,2,3$, are the same up to a shift parameter.
By (\ref{eq:CFY}), we have for $t,s\in\mathcal{R}$, 
\begin{equation}
\label{eq:pf}
\phi_{X_{1}}(t+s)\phi_{X_{2}}(t)\phi_{X_{3}}(s)=\phi_{X_{1}'}(t+s)\phi_{X_{2}'}(t)\phi_{X_{3}'}(s).
\end{equation}
Notice that $\varphi_{j}(t)\equiv\log\phi_{X_{j}}(t)/\phi_{X_{j}'}(t)$
is well defined in a neighborhood of the origin with $\varphi_{j}(0)=0$,
$j=1,2,3$. Thus  for $t$ and $s$ in a neighborhood of zero, \begin{eqnarray*}
\varphi_{1}(t+s)+\varphi_{2}(t)+\varphi_{3}(s) & \equiv & 0.\end{eqnarray*}
 By using the argument of finite differences (c.f. Lemma 1.5.1 of
\cite{kagan:1973}), each $\varphi_{j}$ is a linear complex function
in a neighborhood of zero and thus in $\mathcal{R}$ with the given condition.
That is, there exist complex
numbers $a_{j},b_{j}$ such that
$\phi_{X_{j}}(t)=\phi_{X_{j}'}(t)e^{a_{j}+ib_{j}t}$ 
for any $t\in\mathcal{R}$. By evaluating both sides at $t=0$,
$a_{k}=0$.
By taking the first order derivative on both sides 
at zero, $iE[X_{j}]=iE[X_{j}']+ib_{j}$ and thus $b_{j}\in\mathcal{R}$,
due to $X_{j},X_{j}'\in\mathcal{R}$. Hence $X_{j}$ and $X_{j}'+b_{j}$
have the same distribution. Further, $AE[\mathbf{X}]=AE[\mathbf{X}']$ implies $b_{2}=b_{3}=-b_{1}$.
\end{proof}

For network delay tomography, as a generalization of the simple
two-leaf tree model, let $A$ correspond to a routing matrix
derived from a multicast tree (\cite{presti:2002}), where each node,
except for the root and leaves, must have at least two
children. Let us take the four-leaf tree in Figure \ref{fig:4leaf} as
an example of a 
multicast tree, which will be used for simulation purposes later. Let
$X_1,\cdots, X_7$ denote the link delays on the edges from top to
bottom and from left to right in the tree, i.e., the link delay on the
edge with end node $j$ is denoted by $X_j$. Let $Y_1,\cdots, Y_4$
denote the end-to-end delays from the root node 0 to end node 4, 5, 6 and 7,
respectively. Then each element of $\mathbf{Y}=(Y_1,\cdots,Y_4)^T$ is
a partial sum of $\mathbf{X}=(X_1,\cdots,X_7)^T$, for example,
$Y_1=X_1+X_2+X_4$. This can be written in the form of
\eqref{eq:tomo1}, where $A$ is a $4\times 7$ binary matrix and can be derived
from the linear equations. From Lemma \ref{lemma:lemma}, the
distributions of $X_4,X_5$ are determined up to a
shift parameter by the joint distribution of
$(Y_1,Y_2)$, so are the distributions of $X_6,X_7$. Using a bottom-up 
induction on the tree, it follows that the distributions of all
components of $\mathbf{X}$
are determined by that of $\mathbf{Y}$ up to shift ambiguity. The same
arguments leads to the following theorem.

\begin{thm}\label{thm:delayiden}
Let \(A\) be the routing matrix derived from a multicast tree, then
the distribution of $\mathbf{X}$ is identifiable up to shift
ambiguity.
\end{thm}

Theorem \ref{thm:iden} below provides a general identifiability result for the
traffic demand tomography model, where the routing topology is more
general than a multicast tree, as studied in \cite{cao:2000a}.

\begin{thm}\label{thm:iden}
Let \(B\) be the \([I(I+1)/2]\times J\) matrix whose rows consist of
the rows of \(A\) and the component-wise products of each different
pair of rows from \(A\). If \(B\) has full column rank, then the
distributions of \(\mathbf{X}\) are identifiable up to shift ambiguity. The
shift ambiguity satisfies the constraint \(E[\mathbf{Y}]=AE[\mathbf{X}]\).
\end{thm}

\begin{proof}
For the convenience of expression, ignore the shift ambiguity. Let
$A_iA_k$ be the element-wise product of $A_i$ and $A_k$. Notice
that $(A_{i}A_{k})\mathbf{X}$ denotes the common part of
$(A_{i}\mathbf{X},A_{k}\mathbf{X})$, 
i.e. $(Y_{i},Y_{k})$. Since $\{ X_{j}\}$ are mutually independent,
by Lemma 1, the distribution of $(A_{i}A_{k})\mathbf{X}$ is identifiable.
Thus the distribution of each component of $B\mathbf{X}$ is identifiable.
Let $\psi_{k}$ denote the characteristic function of $B_{k}\mathbf{X}$,
where $B_{k}$ is the $k$th row of $B$. Then for $k=1,\cdots,I(I+1)/2$
and for $t$ in a neighborhood of zero, 
\begin{eqnarray*}
\log\psi_{k}(t) & = & \sum_{j}B_{kj}\log\phi_{X_{j}}(t),\end{eqnarray*}
where $B_{kj}\in\{0,1\}$ is the $(k,j)$th element of $B$. Since
$B$ has full column rank, $\{\log\phi_{X_{j}}(t):j=1,\cdots,J\}$
can be uniquely solved from the above linear equations. 
Then under either (C1) or (C2), $\phi_{X_j}$ is uniquely decided.
That is, the
distribution of each $X_{j}$ can be uniquely identified.
\end{proof}



We now discuss three issues related to the above identifiability results.

{\em 1) Location ambiguity}. 
The location ambiguity of the tomography problem has been recognized
in previous works. To avoid such ambiguity, \cite{vardi:1996}
assumed a Poisson distribution whose mean is the same as its variance,
\cite{cao:2000a} used a power relation between mean and
variance, and all previous discrete link delay models assume
probabilities starting from zero delay. The important message here is
that, despite the location ambiguity, Theorem 1 and 2 state that the
distributional shape of each $X_j$ can be determined, for example, all
orders of central moments that exist are uniquely identified. In
practice, to completely identify the distribution including the
location, one can bring in some additional information such as the
achievable lower bounds of $\mathbf{X}$ for example in delay
tomography, and relationship between mean and variance for example in
traffic demand estimation.

{\em 2) Conditions on the $\mathbf{X}$ distribution}. 
The distributional assumption on $\mathbf{X}$ is very
weak. A lot of well known distributions have analytic characteristic
functions, such as Poisson, Gaussian and discrete distributions, which
have been used in the literature.  A 
mixture distribution that we later use to model the link delays in
Section 4 has an analytic characteristic function. Although, the
heavy-tailed distributions do not 
satisfy (C1), some heavy-tailed distributions such as $\alpha$-stable
distributions satisfy (C2). Despite the generality of our conditions,
we do note that they are not necessary ones.  For theoretical
interest, we have constructed a counter example of a distribution
$\mathbf{X}=(X_1,X_2,X_3)$ that cannot be identified from
$\mathbf{Y}=(Y_1,Y_2)$ for the simple two-leaf tree model $Y_1=X_1+X_2$
and $Y_2=X_1+X_3$, as in the Appendix. 

{\em 3) Condition on the routing matrix $A$}.  \cite{cao:2000a} has
shown that the full rank condition
in Theorem \ref{thm:iden} is necessary in the context of traffic
demand tomography when $\mathbf{X}$ is Gaussian.
In practice, such a condition is easily
satisfied for routing matrices derived from realistic network
topologies.  A more general condition of $A$ has been developed to prove
the identifiability for Poisson distributions in the context of
traffic demand estimation \cite{vardi:1996}. We conjecture that under
Vardi's more general condition of $A$, the distribution of
$\mathbf{X}$ is identifiable up to mean and variance ambiguity under
condition (C1), but we leave the investigation for future work.

\section{Network delay tomography using mixture modeling} \label{sec:mixture}
Below we focus on network delay tomography and describe a
class of flexible mixture models for modeling link delays.
It is well known that there does not exist a standard parametric model
that can sufficiently 
model the distributions of network link delays (see
\cite{duffield:2001b} and \cite{tsang:2003} among others). But it is
possible to define a mixture model which is flexible enough for link
delay distributions.

Assume that for each link $j$, the link delay $X_j$ follows a mixture density
function with $n_j$ components, $X_j \sim f_{X_j}$, defined by :
\begin{eqnarray} 
f_{X_j}(x;\theta_j)  =  \sum_{l=1}^{n_j} p_{jl}\kappa_{jl}(x), \ \ x>0\label{eq:mixture}
\end{eqnarray}
where $\theta_j\equiv (p_{jl},\cdots,p_{jn_j})^T$ contains the mixing
probabilities 
with constraint $p_{jl}\geq 0$, $\sum_{l}p_{jl}=1$, and
$\{\kappa_{jl}\}$ are some basis density functions. 
There is another practical reason that we use a mixture model for
link delays: 
The characteristic function of a mixture distribution is a mixture of
characteristic functions of the basis distributions and thus can be
computed conveniently once the basis distributions are chosen
appropriately, as shown later. In 
this case, the characteristic function of $X_j$ 
can be expressed as 
\begin{eqnarray} \phi_{X_j}(t;\theta_j) & = &
\sum_{l=1}^{n_j}p_{jl}\phi_{jl}(t),
\label{eq:cf-mixture}
\end{eqnarray}
where $\phi_{jl}$ is the characteristic function of the basis function
$\kappa_{jl}$. 

The basis functions are chosen as follows for modeling link
delays. For $j=1,\cdots,J$, let
$0=b_{j1}<b_{j2}<\ldots<b_{j(n_j-1)}<\infty$. Define the basis
function as
\begin{eqnarray}\label{eq:kernel}
\left\{
\begin{array}{ll}
\kappa_{j1}(x) \ = & \mbox{point mass at zero (for zero)}\\
\kappa_{jk}(x) \ = & \mbox{uniform on $[b_{j(k-1)}, b_{jk}]$} \\
  & \hspace*{0in} \mbox{$\ 2 \leq k \leq n_j-1$ (for body)}\\
\kappa_{jn_{j}}(x) \ = & \mbox{exponential with scale $\alpha_j$}\\
  & \hspace*{0.1in}\mbox{on $[b_{j(n_j-1)}, \infty]$ (for tail)}\\
\end{array}
\right.
\end{eqnarray}
The point mass at zero link delay is used here because it is well
known that for a FIFO queue (First In, First Out), the steady state
queuing distribution has 
zero delay with probability one minus the utilization of the queue.
For the body of the distribution, we choose the piecewise uniform
model because of its simplicity and flexibility. Finally, an
exponential distribution is used to model the tail because it is the
right model for the short range dependent traffic model, and for
long-range dependent model it represents a trade-off between accuracy
and simplicity. 
 
In order to reduce the computational complexity, we choose the bin
endpoints $\{b_{jk}\}$ in advance. The parameter of interest is
composed of the mixture coefficients, denoted as
$\theta={\theta_1,\cdots,\theta_J}$. To our advantage, we do not
require 
the bins to be equally spaced. In fact, it is important to choose the
bins that are adaptive to individual link delay distributions in order
to obtain accurate estimates.  Such a varying bin strategy is
especially important for a heterogeneous network environment whose
link delay distributions vary widely across links, because a single
bin width value could be at the same time too coarse grained for a
high bandwidth link with small delays but too fine-grained to
efficiently capture the essential characteristics of the delay along a
low bandwidth link (\cite{presti:2002}). In addition, since a typical
delay distribution may have a density varying a lot at different
areas, it is important to be able to place more bins in the high
density area and fewer bins in the low density area.

\begin{figure}[h]\begin{center}
\psfig{figure=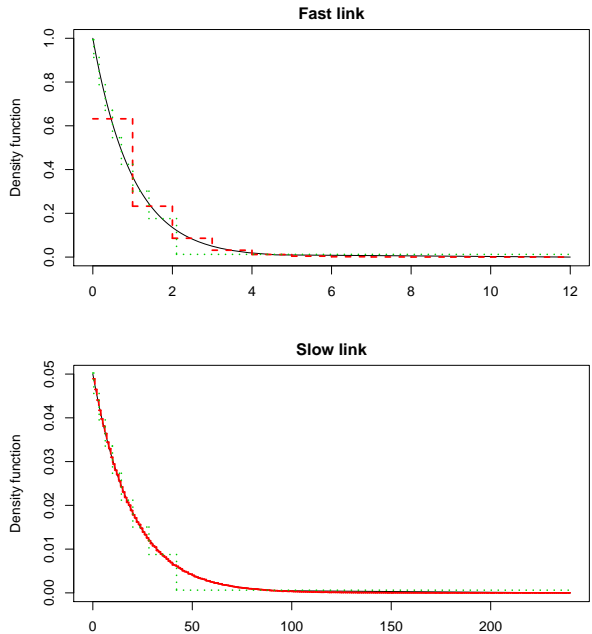,width=3in}
\caption{Fitting two link delay distributions using an equal-bin 
  model and a varying-bin model, where the delay on the slow link
  (bottom) is 20 times in average of that on the fast link (top): the
  solid lines are for the link delay density functions, slashed lines
  for the estimated densities using bin-with equal to 1 (12 bins for the fast
  link and 240 bins for the slow one), and the dotted lines for the
  estimated densities using varying bin-widths (only 10 bins for each).}
\label{fig:delaypdf}\end{center}
\end{figure}

Note that the equal-bin distribution used by most previous
researchers is also a mixture model. Figure \ref{fig:delaypdf} shows
two link delay distributions (in solid lines) where one ranges from
0 to 12 (top) and the other from 0 to 240 (bottom). The slashed lines
are fitted 
curves using a equal-bin model with bin-width 1 which accomodates the
scales of both links, but with 12 bins for the slow link and as many
as 240 bins for
the fast link. The dotted lines are fitted curves using
varying-bin models where only 10 bins are used for both links. It is
clear that the equal-bin model fits the slow link very well, but not
the fast link, while the varying-bin model with a small number of bins
fits both links well. It is possible to use a very small bin-width for
the equal-bin model, but it would require too many bins for the slow
link. The varying bins here are chosen based on quantiles of the delay
distribution, which works very well in general from our
simulation experiences. 
In reality the quantiles are unknown and
we can only obtain an approximation using an initial estimate of the
link delay distribution. This process can be iterated until we get a
good estimate. 

The scale parameter $\alpha_j$ for the tail basis in
Equation~(\ref{eq:kernel}) is unknown and needs to be estimated.  
However, the accuracy of the scale estimate is less important if the
endpoint $b_{j(n_j-1)}$ of the last bin can be placed at the far end of the
tail. For a further simplification, we can fix the tail basis
with a crude estimate of $\alpha_j$ for each link $j$ and  only
estimate the mixing probabilities $\{p_{jk}\}$, which is described next. 


\section{Fast algorithms derived from the General Method of
  Moments}\label{sec:algorithm} 

In this section, we discuss how to estimate the
parameters of mixture coefficients. It is worth pointing out that by
Theorem \ref{thm:delayiden} the parameter of the link delay mixture models 
defined above is identifiable when link delays have positive probabilities 
at zero, which is usually true.

\subsection{The General Method of Moments}\label{sec:estimation}

Following \cite{bu:2002}, it is not hard to show that the
computational complexity of MLE using an EM algorithm for the
above flexible mixture model is of order $O(\max_jn_j^J)$, which is
too expensive.
In this section, we present an estimation approach for network
tomography using Fourier transform, following the pioneering work of
\cite{feuerverger:1977}. The estimators using this approach can be
computed easily as shown below and also exhibit good statistical properties.
The motivation is that the characteristic function of $\mathbf{Y}$ is
simply the {\em product} of the characteristic function of components
of $\mathbf{X}$ as shown in Equation (\ref{eq:CFY}), though the
distribution function is a high order convolution of those of $X_j$s.
We derive the estimator from the General Method of Moments formally described
in \cite{hansen:1982} and \cite{carrasco.florens.2000}. To be
self-contained, we give a formal description of our estimator for the
tomography model below and discuss its advantages over previous
approaches.

Suppose that each $X_{j}$ is modeled by a probability density function
$f_{X_{j}}(x_j;\theta_{j})$ with an unknown parameter $\theta_j$. 
Let $\theta=\{\theta_j: j=1,\ldots,J\}$.
By Equation~(\ref{eq:CFY}), the joint characteristic function of
$\mathbf{Y}$ is 
\[
\phi_{\mathbf{Y}}(\mathbf{t};\theta)=\prod_{j=1}^J\phi_{X_j}(t_j;\theta_j)
\] 
where $\phi_{X_j}$ is the characteristic function with respect to $f_{X_j}$.
Let $\{ \mathbf{Y}(n):1\leq n\leq N\}$ be the independent measurements of $\mathbf{Y}$. The empirical characteristic function of $\mathbf{Y}$ is
\[
\hat{\phi}_{\mathbf{Y}}(\mathbf{t})=\frac{1}{N}\sum_{n=1}^{N}\exp(i\mathbf{t}^{T}\mathbf{Y}(n)).
\]
Similar to the maximum likelihood estimate which is derived by
minimizing the Kullback-Leibler divergence between the empirical
distribution and the model distribution of $\mathbf{Y}$, an estimate
of $\theta$ can
be obtained
by minimizing an $L_2$ distance between the empirical
characteristic function and the model characteristic function of
$\mathbf{Y}$, i.e., 
\begin{eqnarray} \hat{\theta} & = &
\arg\min\int\left|\epsilon_{N}(\mathbf{t};\theta)\right|^{2}d\mu(\mathbf{t}),\label{eq:estimator}\end{eqnarray}
where
\[
\epsilon_{N}(\mathbf{t};\theta)=\sqrt{N}(\hat{\phi}_{\mathbf{Y}}(\mathbf{t})-\phi_{\mathbf{Y}}(\mathbf{t};\theta)),
\]
and $\mu(\mathbf{t})$ is a specified probability distribution
function on $\mathcal{R}^{I}$ (we use the sub script $N$ to show the
dependence on the sample size $N$).  

For a continuous measure $\mu$, the right hand side of
(\ref{eq:estimator}) does not have a closed form in general. To
evaluate the integral, a Monte Carlo approximation can be used: 
first randomly draw $K$ samples from $\mu(\mathbf{t})$, say 
$\{\mathbf{t}_{k}:k=1,2,\cdots, K\}$, and then replace $\mu(\mathbf{t})$ by
its empirical distribution based on these samples.  

Let $\epsilon_{N}(\theta)\equiv(\epsilon_{N}(\mathbf{t}_{k};\theta), k=1,\cdots,K)^T$
be a column vector. We can rewrite (\ref{eq:estimator}) as
\begin{eqnarray} \label{eq:CF}
\hat{\theta}  = 
\arg\min_{\theta}\epsilon_{N}^{T}(\theta)\epsilon_{N}^{*}(\theta),
\end{eqnarray}
where $\epsilon_{N}^*(\theta)$ is the conjugate of
$\epsilon_{N}(\theta)$.  We call it the {\em CF-estimator}, since it
is based on characteristic function. 

The CF-estimator can be considered as a least square estimator based
on the residuals evaluated at $\mathbf{t}_1,\ldots,\mathbf{t}_K$,
which are obviously correlated. Let $W$ be the covariance matrix of $\epsilon_N(\theta)$, it is easy to show that 
\[
W_{jk} 
=\phi_{\mathbf{Y}}(\mathbf{t}_{j}-\mathbf{t}_{k};\theta)-\phi_{\mathbf{Y}}(\mathbf{t}_{j};\theta)\phi_{\mathbf{Y}}^{*}(\mathbf{t}_{k};\theta).
\]
This motivates a weighted version of the CF-estimator, called {\em WCF},
\begin{eqnarray} 
\hat{\theta}^{(W)} & = &
\arg\min_{\theta}\varepsilon_{N}^{T}(\theta)(W+\delta_N I_{K})^{-1}\varepsilon_{N}^{*}(\theta),\label{eq:WCF}
\end{eqnarray}
where $I_{K}$ is the $K\times K$ identity matrix and $\delta_N$, a
tuning parameter, is used to make sure the inversion is well defined.
$\delta$ should be small and we typically choose $\delta_N$ of order
$N^{-1/2}$.  In practice $W$ cannot be calculated precisely since
$\theta$ is unknown.  We can either use a $W$ estimated from an
initial estimate of $\theta$ such as the CF-estimator, or iterate this
process using an iteratively reweighed least squares, which is a
common technique used in generalized linear models.

{\em 1) Statistical properties}.
The characteristic-function based estimators presented in this section
fall into the class of Generalized Methods of Moment (GMM)
estimators. There is a considerable body of work on their statistical
properties, see \cite{feuerverger:1977} and 
\cite{carrasco.florens.2000}, from which the consistency and
asymptotic normality of 
both CF-estimator \eqref{eq:CF} and WCF-estimator \eqref{eq:WCF}, can
be established. In addition, it has been 
proved by \cite{carrasco.florens.2000} that when the probability
measure $\mu$ in 
Equation~(\ref{eq:estimator}) has a density all over $\mathcal{R}^{I}$, the
WCF estimator is asymptotically as efficient as MLE with $K=\infty$
and an appropriate choice of $\delta_N$.

{\em 2) Sampling of $\mathbf{t}$}. For both the CF and WCF estimators, the
points $\mathbf{t}_k,k=1,\ldots,K$ are sampled based on a probability
measure $\mu$. In general, how to choose $\mu$ or sample $\mathbf{t}$
efficiently is a hard problem (\cite{feuerverger:1977}).
In the following, we suggest the choice of $\mu$ based
on our simulation experiences.  

Since the scales of components of $\mathbf{Y}$ may be different,
we normalize $\mathbf{Y}$ by its empirical covariance
matrix and use an elliptic distribution for $\mu$, such as Gaussian.
From simulations we notice that sampling $\mathbf{t}$
directly from a probability measure in $\mathcal{R}^I$ does not easily
yield good results. This is due to the sparsity of $\mathbf{t}$ in the
high dimensional space so that the characteristic functions
$\phi_{\mathbf{Y}}(\mathbf{t})$ evaluated at most of the points are
close to zero. Since the variance of the residual
$\epsilon_N(\mathbf{t};\theta)$ is equal to
$1-|\phi_{\mathbf{Y}}(\mathbf{t})|^2$, the closer to zero of the
characteristic function $|\phi_{\mathbf{Y}}(\mathbf{t})|^2$, the
larger the variance, and the less the information. 
Although it may lose some efficiency,  simulations suggest that better
performance can be achieved by sampling 
$\mathbf{t}$ from lower dimension subspace,
for example 2-dim 
subspaces.  When we draw $\mathbf{t}$ from a lower dimensional
subspace, it implies that we minimize the residuals (difference
between model and empirical characteristic function) only for these
subspaces. This may be viewed as a counterpart of the pseudo likelihood
approach for network tomography proposed by \cite{liang:2003} but in
the Fourier domain. 

\subsection{A Fast Algorithm by Quadratic Programming} \label{sec:fastalg}
With the mixture model described in Section \ref{sec:mixture}, the
unknown parameter for the model is
$\theta=\{\theta_{j}:j=1,\cdots,J\}$, the mixture coefficients. Now we
describe how to estimate $\theta$ 
iteratively using the approach developed in Section
\ref{sec:estimation}. 

By Equation (\ref{eq:CFY}) and (\ref{eq:cf-mixture}), the model
characteristic function of $\mathbf{Y}$ is
\[
\phi_{\mathbf{Y}}(\mathbf{t};\theta) =
\prod_{j=1}^J\theta_{j}^{T}\Phi_{j}(\mathbf{t}^TA^j),
\]
where  
$\Phi_j(t)=(\phi_{j1}(t),\cdots,\phi_{jn_j}(t))^T$.
The objective function in obtaining the CF estimate defined in
Equation (\ref{eq:CF}) can then be written as 
\[
\sum_{k=1}^K|\epsilon_N(\mathbf{t}_k;\theta)|^2 = 
\sum_{k=1}^K\left|\hat{\phi}_{\mathbf{Y}}(\mathbf{t}_k)-\prod_{j=1}^{J}\left(\theta_j^{T}\Phi_{j}(\mathbf{t}_k^TA^j)\right)\right|^{2}.
\]
It is easy to see that for each $\theta_j$, if the rest of the
parameters are known, the optimization function is a quadratic
function of $\theta_j$. Specifically, given all other parameters
$\{\theta_l:l=1,\cdots,J,l\neq j\}$, the optimal $\theta_j$
can be obtained by minimizing 
\begin{eqnarray} \label{eq:quadprog}
C(\theta_j)
& = &
\theta_j^{T}D_j\theta_j-2\theta_j^{T}\mathbf{d}_j,
\end{eqnarray}
where \begin{eqnarray*} 
D_j & = & \sum_{k=1}^K \left|\prod_{l\neq
j}\phi_{X_{l}}(\mathbf{t}_k^{T}A^{l})\right|^{2}Re\{\Phi_{j}^{*}(\mathbf{t}_k^TA^j)\Phi_{j}^T(\mathbf{t}_k^TA^j)\}
\end{eqnarray*} is an $n_{j}\times n_{j}$ matrix, and
\begin{eqnarray*} \mathbf{d}_j & = & \sum_{k=1}^K
Re\{\hat{\phi}_{\mathbf{Y}}^{*}(\mathbf{t}_k)\prod_{l\neq
j}\phi_{X_{l}}(\mathbf{t}_k^{T}A^{l})\Phi_{j}(\mathbf{t}_k^TA^j)\}
\end{eqnarray*} 
is an $n_{j}$-dim column vector. 
This is a standard quadratic programming problem and can be solved
quickly.
Therefore, estimation of $\theta$ can be obtained by an iterative algorithm as follows.
\begin{alg} Iterative quadratic programming\label{alg:quadprog}
\begin{itemize}
\item[(1)] Choose an initial value for $\theta_j$, $j=1,\ldots, J$.
\item[(2)] For each $j=1,\ldots, J$, 
  estimate $\theta_j$ by minimizing (\ref{eq:quadprog})
  using quadratic programming. 
\item[(3)] Repeat Step 2 until convergence.
\end{itemize}
\end{alg}

A nice property of Algorithm \ref{alg:quadprog} is that it
always converges to a local solution
because the objective function never increases after each iteration
and is bounded below by 0. This is similar to EM algorithms, but care
is needed in order to obtain the global minimal. Simulations show that
$\{ p_{jk}=1/n_{j}\}$ can serve as a good starting value. 

The computational complexity of each iteration in Algorithm
\ref{alg:quadprog} is  
$O(KIJ\max_j(n_{j})^{3})$ for the {\em CF-estimator}. For the {\em
  WCF-estimator}, a similar iterative algorithm by quadratic
programming can be obtained. Due to the weight matrix, the complexity
of each iteration becomes $O(K^3IJ\max_j(n_j)^3)$.  


\begin{figure}[h]\begin{center}
\psfig{figure=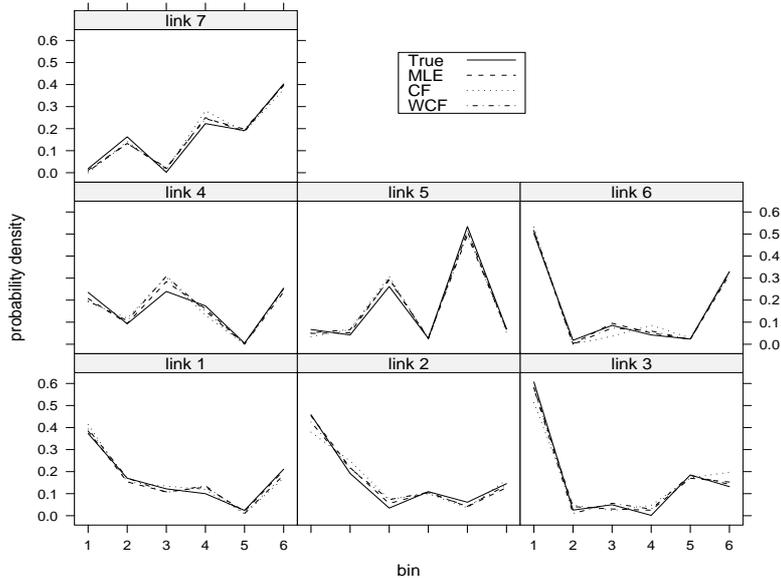,width=4in,height=3in}
\caption{Estimated link delay probability densities on a four-leaf tree
  from 500 end-to-end delay measurements. The true delay distribution
  for each individual link has a discrete probability density with 6
  equally spaced bins.}  
\label{fig:mn_pdf}\end{center}
\end{figure}

\begin{figure}[hbt]\begin{center}
\psfig{figure=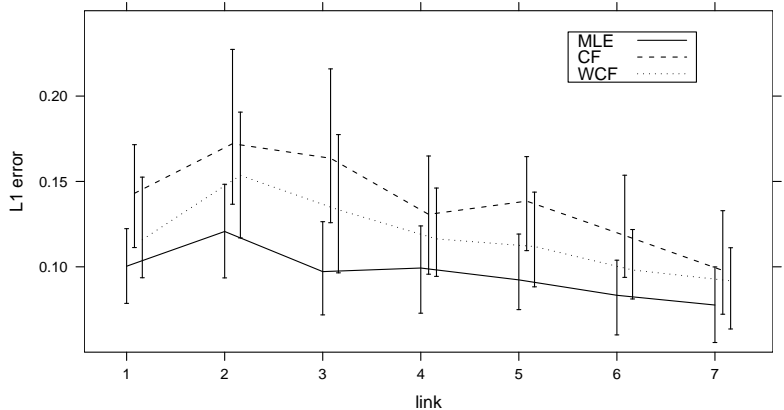,width=4in}
\caption{Quartiles of the $L_1$ error for the three link delay
  distribution estimates from 100 simulation runs, where each
  simulation run has the same setup as in Figure \ref{fig:mn_pdf}. 
The 25\%, 75\% quantiles are shown in line segments, and the median
  are shown in lines. } 
\label{fig:mnL1err}\end{center}
\end{figure}

\begin{figure}[h]\begin{center}
\psfig{figure=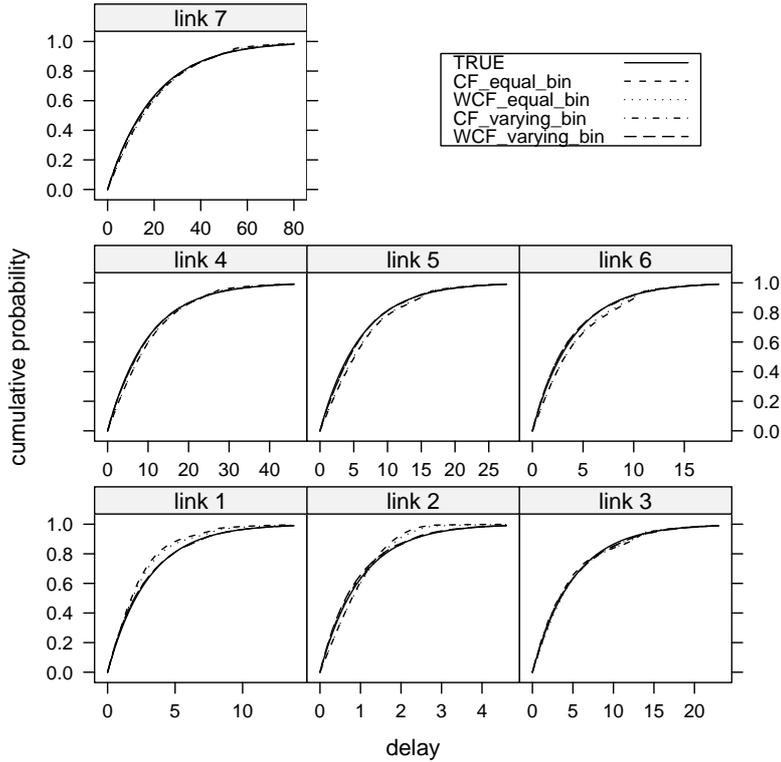,width=4in}
\caption{The cumulative probability of the estimated link delay
  distributions on a four-leaf tree (Figure~\ref{fig:4leaf}) from 2000
  end-to-end delay measurements.  The true delay distributions, shown
  in solid line, are exponentials with means 3,1,5,10,6,4 and 20 for
  link 1 to link 7 on the four-leaf tree respectively (where the link
  index is ordered from top to bottom, left to right). Estimates are
  obtained using a mixture model of piecewise uniform of 12 bins and
  an exponential tail. 
}
\label{fig:expcdf}\end{center}
\end{figure}

\begin{figure}[h]\begin{center}
\psfig{figure=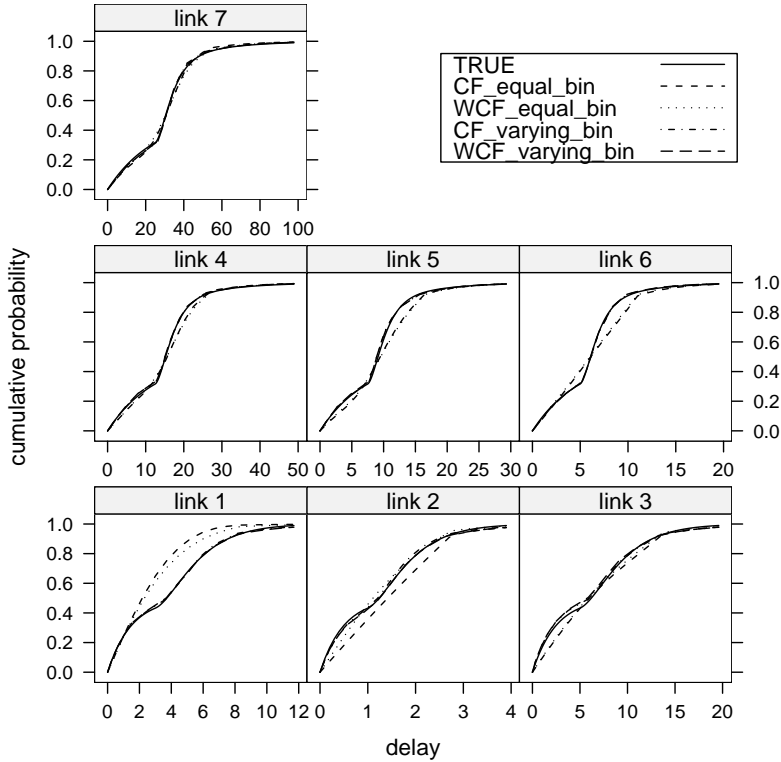,width=4in}
\caption{The cumulative probability of the estimated link delay
  distributions on a four-leaf tree (Figure~\ref{fig:4leaf}) from 2000
  end-to-end delay measurements.  The true delay distributions, shown
  in solid line, are a mixture of exponential and Gamma distribution
  with a shape parameter of 2 with means 3,1,5,10,6,4 and 20 for link
  1 to link 7 on the four-leaf tree respectively (where the link index is
  ordered from top to bottom, left to right). Estimates are obtained
  using a mixture model of piecewise uniform of 12 bins and an
  exponential tail.  
}
\label{fig:expgamcdf}\end{center}
\end{figure}

\begin{figure}[h]\begin{center}
\psfig{figure=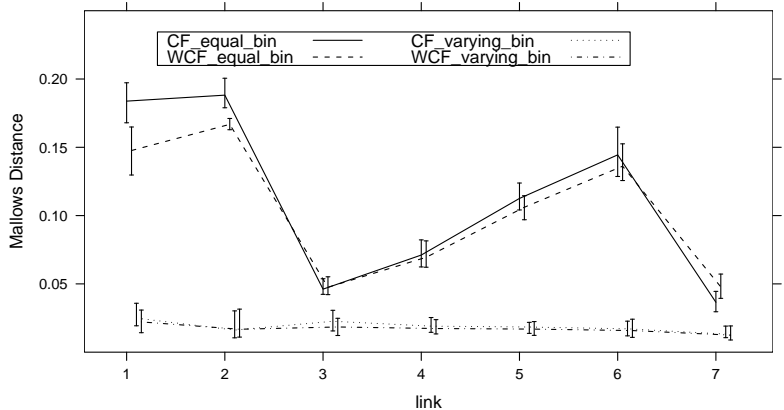,width=4in}
\caption{Quartiles of the normalized Mallows distance for the four
  link delay distribution estimates from 100 simulation runs, where
  the true delay distribution is an exponential. Each simulation run
  the same setup as in Figure~\ref{fig:expcdf}. The first and third
  quartiles are shown in line segments, and the median are shown in
  lines.}
\label{fig:expmallow}\end{center}
\end{figure}

\begin{figure}[h]\begin{center}
\psfig{figure=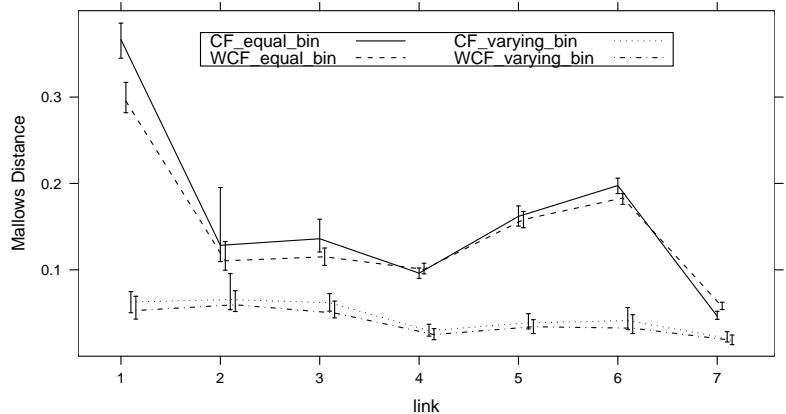,width=4in}
\caption{Quartiles of the normalized Mallows distance for the four
  link delay distribution estimates from 100 simulation runs, where
  the simulated delay distributions are a mixture of an exponential
  and a Gamma with shape parameter 2. Each simulation run the same
  setup as in Figure~\ref{fig:expgamcdf}. The first and third quartiles are
  shown in line segments, and the median are shown in lines.} 
\label{fig:expgammallow}\end{center}
\end{figure}

\begin{figure}[h]\begin{center}
\psfig{figure=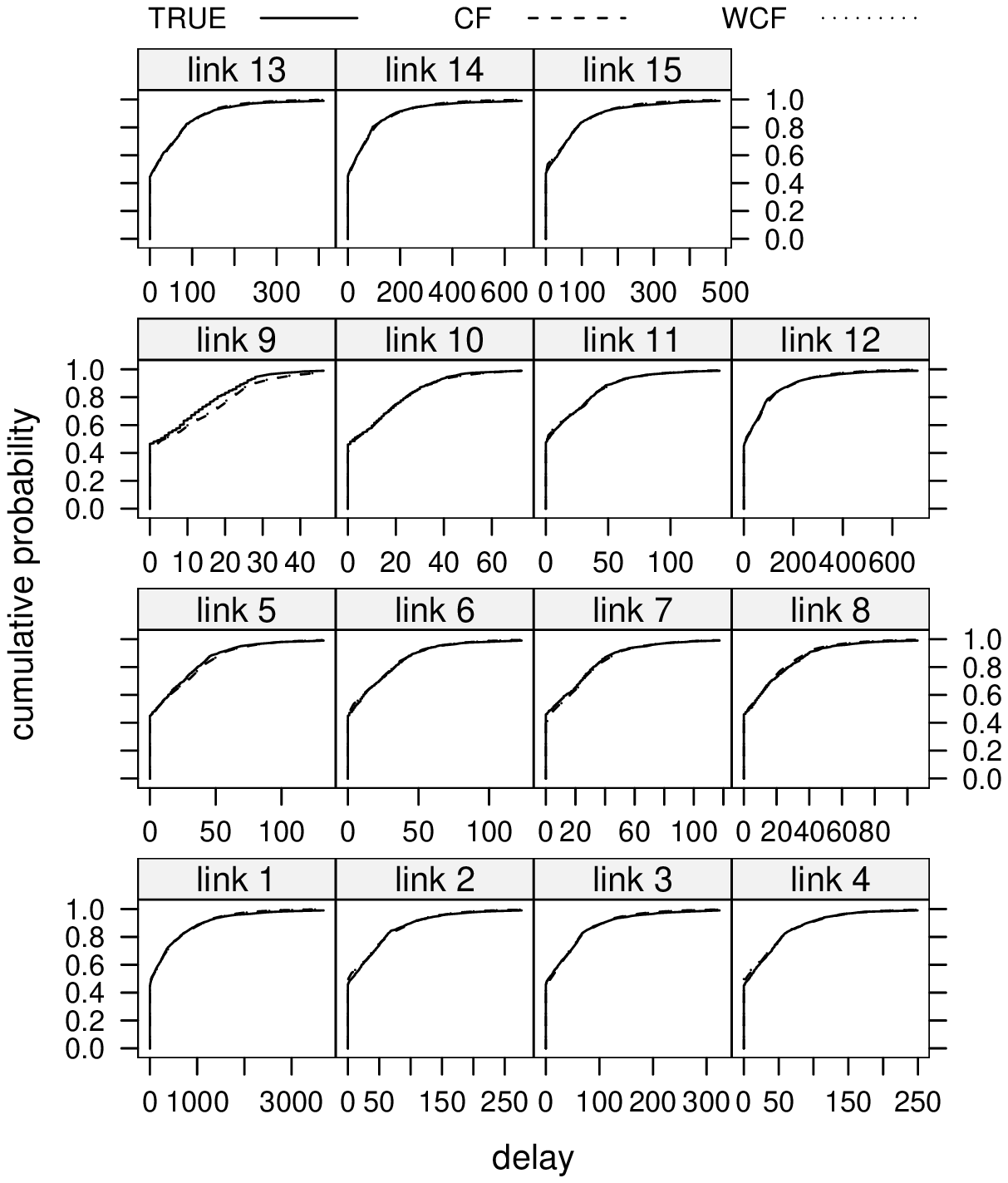,width=3.2in}
\caption{The cumulative probability of the estimated link delay
  distributions on a 8-leaf binary tree from 1800 end-to-end delay
  measurements.  The true delay distributions, shown in solid line,
  have an approximate Weibull distribution excluding the zero delay.
  The average link delays vary by a factor of forty.  Estimates are
  obtained using a mixture model that consists of a point mass at 0, a
  piecewise uniform distribution of 6 bins, and an exponential tail.} 
\label{fig:8leaf_cdf}\end{center}
\end{figure}

\section{Simulation and Experimental Studies}
\label{sec:simulation}
In Section \ref{sec:algorithm}, we have developed simple and fast
algorithms using a flexible mixture model for network delay tomography.
In this section, we evaluate the performance of
the proposed algorithms in terms of statistical efficiency and
accuracy. 
To measure the accuracy of the estimation as compared to
the true distributions, we use a $L_1$-distance for discrete link
delay distributions and a normalized Mallows distance for continuous
link delay distributions. 

Our evaluation is divided into three pieces. 
First, we study the efficiency of our estimates by comparing them
with that of MLE for a discrete link delay distribution with equally
spaced bins. We show that our estimators have comparable efficiency to
that of MLE which is statistically efficient and also computable in
this setting.
Second, we examine the performance of our estimators using model
simulations for continuous link delays in an ideal scenario where both
temporal and spatial independence hold. Model simulations
demonstrate the importance of varying bins selection that should adapt
to not only delay distributions of individual links but the different
scales of delays across links in order to achieve satisfactory estimates.
Finally, we use real trace driven simulations to
examine the accuracy of our estimators under more realistic scenarios
where the independence assumptions may not be strictly true as
appeared in the Internet. Results from our trace driven simulations
demonstrate that the estimates made by our algorithms closely match
the real distributions. 

\subsection{Efficiency Evaluation}
We study the efficiency of our estimators using a discrete link delay
distribution with equally spaced bins on a four-leaf tree
(Figure~\ref{fig:4leaf}).  For link $j, j=1,\ldots,7$, the link delay
has a discrete distribution at $\{0,1,\cdots,5\}$ with probabilities
generated uniformly from the space $\sum_{k=1}^6 p_{jk}=1$
with constraints $0<p_{jk}<1$.  A total of 500 delay samples
are generated for each link from its specified delay distribution and
the end-to-end delays are computed according to the model
\eqref{eq:tomo1}. The delay distributions of all 
seven links are estimated using the MLE, the CF-estimator, and the
WCF-estimator. 

We repeat the experiment 100 times with different random seeds. Both
the MLE and the CF-estimator use the uniform distribution as 
starting values whereas WCF uses the CF estimates as starting
values. The weight matrix $W$ for WCF is also derived from the CF
estimates. For both the CF and WCF estimators, a total of 3000 samples of
$\mathbf{t}$ are drawn randomly from the 2-dim subspaces of $I$-dimensional
end-to-end delays using a Gaussian distribution with a scale parameter of
5 after normalizing $\mathbf{Y}$. We have
also run the recursive algorithm developed in \cite{presti:2002}, but
we do not report the result here except to state that it often yields
much poorer estimates (similar to observations made by
\cite{liang:2003}). 

Figure~\ref{fig:mn_pdf} shows both the estimated and the simulated
seven link delay density functions in one simulation
experiment. We observe that all methods give reasonably accurate estimates. To
compare errors of the different estimators, we calculate the $L_{1}$
distances between the estimated link delay density functions and the
ground truth for each of the 100 experiments. Figure~\ref{fig:mnL1err} reports
the 25\%, 75\% quantiles of the $L_{1}$ errors for each link in
vertical line segments, whose middle points represent median errors.
MLE has the smallest median error, and the median errors of CF and WCF
are 50\% and 22\% higher than that of MLE.
The results suggest that both CF and WCF are somewhat worse than as
expected but comparable to MLE.

\subsection{Accuracy Evaluation Using Model Simulation}
In this subsection, we investigate the performance of the estimators
for continuous delay distributions, which are more realistic than
discrete ones since network delays are essentially continuous except at zero. 
Delay tomography in a heterogeneous network is intrinsically more
challenging than in a homogeneous network because links
with small delays are not equally represented as links with large
delays in the end-to-end delay measurements. In addition, the
heterogeneous environment also represents a situation where most of
the existing methods such as MLE do not work well because they rely on
simple discretization. After all, the real Internet is a heterogeneous
network. Thus we report model simulations on a four-leaf tree that resemble a
heterogeneous network environment. For simplicity, we do not consider
the point mass at zero for model simulations, but we will treat this
in later real trace driven simulations.
 
To conduct a comprehensive evaluation, we run simulations for the link
delay distributions of different shapes. Due to the space limit, we
only report the results for two representative distributions that are
i) exponential (uni-modal) and ii) a mixture of an exponential and a
Gamma with shape parameter 2 
(multi-modal). In
both cases, the average link delays on the four-leaf tree are 3, 1, 5,
10, 6, 4 and 20 respectively for link 1 to 7 assigned from top to
bottom and left to right,  which resembles a heterogeneous network
with the average link delays varying by a factor of 20.  We generate
2000 delay samples for each link from the specified delay
distributions, and we estimate the seven link delay
density functions from the resulting end-to-end delays. 
 
We use four different estimates of link delay distributions: {\em
CF\_equal\_bin, WCF\_equal\_bin, CF\_varying\_bin,
WCF\_varying\_bin}. All four estimates are obtained using a mixture
model for the link delays of the same form as \eqref{eq:kernel} with
$n_j=12$ except removing the point mass at 0.
The difference in the mixture model
for the estimates lies in the bin placement.  For both {\em
CF\_equal\_bin} and {\em WCF\_equal\_bin}, the 12 bins are equally
spaced using a bin width selected for each link based on variance
estimates, which are 
obtained by solving systems of linear equations, following
\cite{duffield:2004}. For both {\em CF\_varying\_bin} and {\em
WCF\_varying\_bin}, the bins are located at the quantiles of
the delay distributions that corresponds to probabilities $i/13,
i=1,\ldots, 12$.

Figure \ref{fig:expcdf} and \ref{fig:expgamcdf} plot the estimated
cumulative distribution functions for each link delay for case i) and
ii) respectively, along with the ground truth in one simulation run.
From the figures, we observe that the estimates
using varying bins are almost identical to the true distributions. The
estimates using equal bins give satisfactory
estimates for case i) but not quite as good for the more complex case ii).

To measure the accuracy of the estimates, we use the  Mallows
distance defined for a cumulative distribution $F$ and its estimate 
$\hat{F}$ by
\[
M(F,\hat{F}) = \int_{0}^1 \left|F^{-1}(p) - \hat{F}^{-1}(p)\right| dp,
\]
where $F^{-1}$ and $\hat{F}^{-1}$ are the inverse cumulative
distributions.  The Mallows distance can be viewed as the average of
absolute difference in quantiles between two distributions. Because
the Mallows distance is linear to the scale of distributions, we use
$M(F,\hat{F})/\sigma_F$, the normalized Mallows distance, to measure
the difference between $F$ and $\hat{F}$, where
$\sigma_F$ is the standard deviation of $F$. 

We repeat the simulation 100 times and compute the normalized Mallows
distance between the estimated and true distributions as the error
metric for all 
links. Figure~\ref{fig:expmallow} and \ref{fig:expgammallow} report,
corresponding to
case i) and ii) respectively, the first and third quartiles of
the errors as well as median errors for each link, similar to Figure
\ref{fig:mnL1err}.
It is clear that the varying bins improve the
quality of estimates significantly over equal bins. (Note that the
difference between CF and WCF are not significant though.)
This suggests that
selecting bins based on characteristics of the underlying density
distributions is important in improve the accuracy in a heterogeneous
network. 

\subsection{Accuracy Evaluation Using Real Internet Traces} 
We next investigate how the algorithms perform in a realistic network
environment
where some of the assumptions may not hold completely.
For instance, due to the closed-loop control
nature of the TCP protocol, the packets within the same TCP connection
have strong temporal dependency. Although the dependency is weakened
when many TCP connections are multiplexed as they arrive to a link,
the dependency may not be completely gone. We approximate a real
scenario by simulating the behavior of a link using the real traces
collected from the Internet. Since the traces include the arrival time
and the size of each packet, the simulation sees the exact link
behaviors as what the original link where the trace collected from
seen if we set the bandwidth and the buffers the same as the original
link. 

We use traces from the NLANR web site
\footnote{http://pma.nlanr.net/Traces/} that archives packet header
traces collected from about ten links at different locations of the
Internet. The links differ in both bandwidth and traffic. A 90-second
trace is recorded every one (two) hours for each of the links. In our
experiment, we first assign traces collected from different sites to
the links of the simulated network. We then simulate the
links using 
the assigned traces as input using the standard network simulator tool
\cite{ns}.  Moreover, we superpose the probes to 
the traces and record their per-link queuing delays as well as
end-to-end delays where the latter is used as input for the estimation
whereas the prior is for comparison with the estimates. 

Notice that the delays on
the edge links in a real network may vary more than the core
links due to its low 
bandwidth. In addition, the average delay may also differ dramatically
for different links. We resemble a real network in a symmetric binary
8-leaf tree by 
assuming that both the root and the leaves of the tree are on the edge
of the network whereas the interior links are in the core. We assign
traces of high rates to links in the core and traces of low rates to
the edge links.   

Figure~\ref{fig:8leaf_cdf} shows both CF and WCF estimate of the delay
distribution using a varying bin strategy laid out in
Section~\ref{sec:mixture}, along with the simulated distribution. The
throughputs across different links vary by a factor of 40. It is
easy to see that the estimates are extremely good for most links,
except for link 9 that has the smallest average link delay where it
shows some marginal error. The average normalized Mallows distance
over all links is 0.065 which also suggests a good match between the
estimates and simulated results. We have also simulated the four-leaf
tree network and a 
symmetric binary 16-leaf tree network respecitvley, using different
traces, and the proposed algorithms give satisfactory results.

In addition, our real trace driven simulations suggest that the link
delay distributions excluding the tail can be well approximated by a
Weibull distribution with a shape parameter slightly smaller than
1. This is not surprising because it has been shown that the queuing
delay for a FIFO queue with a Fractional Brownian Motion traffic input
has a Weibullian tail. The Weibullian form is also consistent with the
finding in \cite{cao:2004}.

\section{Conclusion}
This paper has presented a general identifiability result and
introduced a general estimation approach for the network
tomography problem. For network delay tomography, a fast algorithm
based on GMM has been developed for estimating the link delay
distributions using mixture models of characteristic functions.
In comparison with likelihood based approaches, the most significant
nature of the new method is that it affords the choice of varying
bin widths which adapts to delay variabilities of individual links and
has low computational complexity. The new approach can be applied to
traffic demand estimation as well.

\section*{Acknowledgments}
We would like to thank Gang Liang for sharing his simulation codes and
Michael Greenwald for helpful discussions. A conference version of the
main results has appeared in the Proceeding of IEEE INFOCOM
(\cite{chen.et.al.2007}).

\section*{Appendix: A counter example}

\begin{figure}
\begin{center}
\includegraphics[width=3.6in]{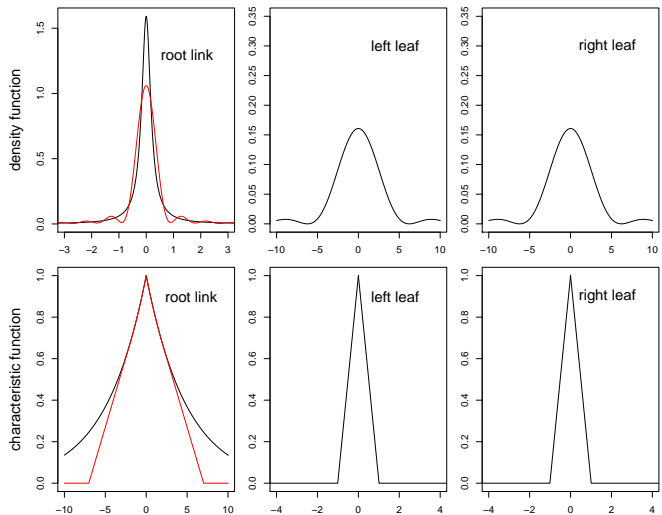}
\end{center}
\caption{A counter example of identifiability for the two-leaf tree
  model where $(X_1+X_2,X_1+X_3)$ and $(X_1'+X_2',X_1'+X_3')$ have the same
  joint distribution: The bottom three figures plot the characteristic
  functions: the first one for $\phi_{X_1}$ and $\phi_{X_1'}$, and
  the second and third ones for
  $\phi_{X_2}=\phi_{X_2'}$ and $\phi_{X_3}=\phi_{X_3'}$
  respectively; The top three figures plot the
  corresponding probability density functions.}
\label{fig:counter}
\end{figure}

Based on the proof of Lemma 1, we can construct a counter example
that the distributions of $\{ X_1,X_2,X_3\}$ are not identifiable. Let $c(t;a,\lambda)=e^{-\lambda|t|}I(|t|\leq a)+\lambda e^{-\lambda a}(a+\frac{1}{\lambda}-|t|)I(a<|t|\leq a+\frac{1}{\lambda})$
be a continuous function defined for $t\in\mathcal{R}$. It is easy
to check using Polya's condition (\cite{lukacs:1970}) that for any $a\geq0,\lambda>0$, $c(t;a,\lambda)$ is a
characteristic function corresponding to a symmetric, non-vanishing,
bounded continuous density function. Let $\phi_{X_1}(t)=c(t;2,1)$,
$\phi_{X'_1}(t)=c(t;3,1)$,
$\phi_{X_2}(t)=\phi_{X'_2}(t)=\phi_{X_3}(t)=\phi_{X'_3}(t)=c(t;0,1)$. 
Both groups of distributions corresponding to characteristic functions
$\{\phi_{X_k}\}$  and $\{\phi_{X'_k}\}$ for $\{ X_{k}\}$ respectively can
generate the same joint
distribution of $(Y_{1},Y_{2})$. Figure \ref{fig:counter} shows both
their characteristic 
functions and probability density functions on a two-leaf tree:
$\phi_{X_1}$ and $\phi_{X_1'}$ are the two curves in the first box of
the second row, and $\phi_{X_2}$ and $\phi_{X_3}$ are in the second
and third box of the second row respectively, with corresponding
density functions plotted in the first row. Notice that
these distributions cannot be used for link delays which do
not permit negative values. It is an open question whether there
exist link delays whose distribution shapes are 
not identifiable even for the simple two-leaf tree tomography.

\bibliographystyle{Chicago}
\bibliography{tomo}

\end{document}